\documentclass[suppldata]{interact}

\usepackage{epstopdf}
\usepackage{threeparttable}
\usepackage{algorithm}
\usepackage{tabularx}
\usepackage[noend]{algpseudocode}
\usepackage{url}
\usepackage[caption=false]{subfig}

\usepackage[longnamesfirst,sort]{natbib}
\bibpunct[, ]{(}{)}{;}{a}{,}{,}
\renewcommand\bibfont{\fontsize{10}{12}\selectfont}

\theoremstyle{plain}

\theoremstyle{definition}

\theoremstyle{remark}

\begin{document}

\title{Outpatient Diversion using Real-time Length-of-Stay Predictions}

\author{
\name{Najiya Fatma\textsuperscript{a}\thanks{CONTACT Varun Ramamohan. Email: varunr@mech.iitd.ac.in} and Varun Ramamohan\textsuperscript{a}}
\affil{\textsuperscript{a}Department of Mechanical Engineering, Indian Institute of Technology Delhi, New Delhi, 110016, India}
}

\date{}

\maketitle

\begin{abstract}
In this work, we show how real-time length-of-stay (LOS) predictions can be used to divert outpatients from their assigned facility to other facilities with lesser congestion. We illustrate the implementation of this diversion mechanism for two primary health centers (PHCs), wherein we divert patients from their assigned PHC to the other PHC based on their predicted LOSs in both facilities. We develop a discrete-event simulation model of patient flow operations at these two PHCs in an Indian district and observe significantly longer LOSs at one of the PHCs due to disparities in the patient loads across both PHCs. In this context, we first determine the expected LOS of the patient at the points in time at which they are expected to arrive at both PHCs using system state information at the PHC recorded at the current time. The real-time LOS predictions are generated by estimating patient wait times on a real-time basis at the queueing subsystems within the PHC. We then divert the patient to the appropriate PHC on the basis of the predicted LOS estimates at both PHCs, and show through simulation that the proposed framework leads to more equitable utilization of resources involved in provision of outpatient care.
\end{abstract}


\begin{table}[h]
	\begin{threeparttable}
		\caption{Key acronyms and notation.}\label{tab:acronyms} \centering
		{\begin{tabular}{|c|c|}
				\hline
				Acronyms & Corresponding constituent words  \\
				\hline
				LOS & Length-of-stay \\
				\hline
				PHC & Primary health center\\
				\hline
				DES & Discrete event simulation \\
				\hline
				$\delta_{x}$ & Travel time to reach PHC $x$\\
				\hline
				$\rho$ & Resource utilization \\
				\hline
				$\Delta_{\rho}$ & Change in resource utilization \\
				\hline
				$w$ & Wait time in a queue \\
				\hline
				$\Delta_{w}$ & Change in queue wait time \\
				\hline
				$G(.)$ & Queueing system service time $cdf$ \\
				\hline
		\end{tabular}}
		\label{tabn}%
	\end{threeparttable}
\end{table}

\section{Introduction}

Patient length of stay (LOS) is an important indicator of the efficiency of hospital management systems \citep{Ayy20}. Long LOSs and overcrowding are important operational issues facing healthcare facilities around the world \citep{wachtel2020addressing}, including in India \citep{sharma2021overcrowding}. This in turn leads to both operational and clinical issues, such as denial of hospital admissions, cancellation of surgeries, and higher likelihoods of developing healthcare acquired infections \citep{zhou2019estimating,arefian2019estimating}.

Congestion and longer LOSs at specific facilities within a healthcare facility network, which in turn lead to inequitable utilization of facilities within the network, may occur due to patient perceptions of better quality of care at said facilities within the network \citep{liu2018patients}. Healthcare administrators may adopt different strategies to relieve congestion such as capacity replanning \citep{whiteside2020redesigning}, patient diversion \citep{marquinez2021identifying}, and early discharges for medically stable patients. However, increasing resource capacity is often difficult due to the capital investment required and the attendant medical training required when personnel numbers are increased. Hence, patient diversion is often the preferred alternative to mitigate overcrowding. In this study, we describe an approach towards using real-time LOS predictions to facilitate diversion of outpatients across primary healthcare facilities. While we demonstrate our approach via the diversion of outpatients across two primary healthcare centers in the Indian context, our approach can be used in any setting where patient diversion is being considered due to high LOSs, or more generally, where significant disparities in resource utilization in a healthcare network is observed.

In India, PHCs are the first point of contact for the patients with a formally trained doctor and provide outpatient care, and limited emergency and inpatient care \citep{shoaib2021simulation}. There are approximately 30,000 PHCs in India, with each PHC mandated to serve the primary care needs of a population of approximately 30,000 persons \citep{phc2020}. The Indian government has recently undertaken an initiative to expand and upgrade existing primary health centers (PHCs) \citep{blanchard2021vision}, and hence it is reasonable to assume an increase in the patient load at these PHCs given the established link between increased quality of care, improved infrastructure and higher demand \citep{rao2018quality}. Based on previous work regarding modelling PHC operations \citep{fatma2020primary,shoaib2021simulation}, we developed a discrete-event simulation (DES) of medical care operations of two PHCs in an Indian district. Under conditions of high outpatient demand, the DES outcomes indicated signifcant disparity in outpatient LOSs across both facilities, and this would likely extend to the entire public health network in the district. Further, as we describe in detail in Section 3, we also noted that under these high demand conditions, patient wait times to receive service from the doctor did not increase significantly due to the very high service rates of the doctor. Instead, the impact of high demand was reflected in wait times at supporting subsystems in the PHC such as the pharmacy and the laboratory, which in turn yielded high LOSs in the PHC as a whole. For example, average LOSs were approximately 60 minutes, in comparison to average total service times of less than 10 minutes across all services that a patient may avail of in a typical PHC visit. Thus, implementing diversion on the basis of wait times in this context did not appear appropriate; hence, we considered diversion based on patient LOS as a whole in the facility.

The diversion mechanism that we propose utilizes real-time LOS predictions to inform the diversion decision. Diversion mechanisms based on real-time wait time or `delay' predictions for emergency patients have been discussed in previous work \citep{fatma2021patient}; however, in this work, we use real-time delay estimates to calculate real-time LOS predictions to determine the diversion decision. Note that these real-time delay estimates, and in turn the real-time LOS predictions, are generated at the point when the patient decides to visit the facility in question (as opposed to when the patient arrives at the facility), implying that an information technology (IT) system, including a web/smartphone application, to convey this information to this patient would be required. Examples of such IT systems can be found in the United Kingdom's National Health Service \citep{mustafee2020providing}. In this paper, we develop a mechanism for patient diversion based on such real-time LOS predictions (including a demonstration of how such LOS predictions can be generated) and demonstrate via DES its impact on the operational outcomes of a healthcare facility network. Our study provides the operational basis that an IT system such as that described in \cite{mustafee2020providing} can leverage to implement real-time LOS-based diversion.

We now discuss the literature relevant to this study, and describe our research contributions with respect to the literature.

\section{Literature Review}

In this section, we focus on the literature associated with real-time LOS prediction and patient diversion. As part of the literature review, we also discuss studies that highlight the importance of information sharing across facilities in the network wherein diversion is implemented. We describe the research contributions of this work at the end of each subsection.

\subsection{Length of Stay Prediction}\label{class}

The importance of LOS prediction at healthcare facilities and their subunits such as the emergency department is reflected in the considerable body of literature devoted to this topic \citep{song2015diseconomies,alsinglawi2020predicting,shi2021timing}. From the perspective of the methods used for LOS prediction, it appears that data mining and statistical learning methods have been most widely used in LOS prediction \citep{verburg2014comparison,barnes2016real,turgeman2017insights,gutierrez2021predicting,shaaban2021statistical}. A review conducted by \cite{el2021machine} found that multiple machine learning techniques such as deep neural networks, random forest trees, regression techniques and decision trees among others have been used to predict the patients LOS. The authors then conclude that the LOS predictions were found to be useful in planning patient admission schedules and, as required, the preparation of post-discharge care.

LOS prediction studies in the literature have primarily focused on inpatients and emergency patients using statistical learning techniques. The typical methodology for LOS prediction for these patient types involved the following steps: (a) recording clinical information of patients such as medication history at the time of admission, patient age, and the number of comorbidities, (b) identifying the set of predictor variables among the data available for LOS prediction, (c) training and validating the statistical learning methods on the dataset constructed via mining the available data. Examples of such studies include \citep{baek2018analysis,aghajani2016determining,hijry2020application}, wherein the authors concluded that factors such as severity of the disease, recency of diagnosis and type of surgery, patient age, number of comorbidities, surgery type, number of days of hospitalisation before surgery, etc., were significantly correlated with patient LOS. These studies did not determine the effect of operational variables such as the number of patients in the queue for system resources, the elapsed service time of patients currently receiving care, etc., on patient LOS. 

In this context, we generate LOS predictions via a methodology grounded in queueing theory and real-time delay prediction that estimates patient LOSs in the PHC as a function of operational `system state' variables such as the number of patients in the queue at each queueing system within the PHC (e.g., the queueing system represented by the doctor and the laboratory), the elapsed service times of patients currently receiving service at each queueing subsystem, arrival and service rates of patients in the PHC, and the average travel time of patients to the PHC from their point of origin. If the patient decides to leave for the PHC at time $t$, then the LOS prediction is generated for the time of arrival of the patient in the system - i.e., at time $t + \delta$, where $\delta$ is the travel time to the PHC from the patient's location.

We note here that because we use real-time delay predictions to generate the LOS predictions, our approach can be extended to multi-class multi-server systems for which real-time delay predictors have previously been developed \citep{ibrahim2018sharing}. Further, in general, our approach requires fewer variables to be recorded for generating the real-time delay estimates in comparison to statistical learning predictors, and its development and deployment will not require construction of a training and validation dataset. We also note that the majority of previous work focuses on generating LOS predictions for patients at a single unit of a healthcare facility, such as the emergency department, and do not generate predictions for patients that receive service from multiple units within a given facility. In comparison, our approach generates LOS predictions at multiple units within the same facility; that is, total predicted LOS is the summation of time spent at each PHC subsystem.

\subsection{Patient Diversion}

There are multiple studies considering patient diversion as a means to reduce congestion at healthcare facilities. In a review by \cite{li2019review}, the authors find that patient diversion was initiated based on multiple measures of congestion, such as the average wait time of patients before receiving care, the number of patients waiting to receive care, the number of occupied beds in a ward, etc. From a methodological standpoint, the approaches used to optimize various aspects of the diversion process include a Markov decision process formulation in \cite{marquinez2021identifying} to maximize the cost-effectiveness of diversion policies for ICU patients in a public-private hospital network, a simulation optimization approach in \cite{ramirez2011design} to minimize the time to receive care (including wait times and travel times between facilities), a mixed integer programming approach in \cite{nezamoddini2016modeling} to minimize the number of patients waiting for care across the network. \cite{fatma2021patient} evaluated via DES the effectiveness of patient diversion strategies on multiple operational outcomes such as resource utilizations, wait times, and the proportion of patients whose wait time before receiving care exceeds a threshold duration. The proportion of patients leaving without receiving care has also been an operational outcomes of interest in multiple studies, as summarized in \citep{morley2018emergency}.

An important consideration in implementing a diversion mechanism within a healthcare facility network is information sharing across the network regarding operational variables at each facility that are used to determine diversion decisions. In this regard, there is considerable support in the literature regarding implementing centralized diversion policies across healthcare facilities, because diverting patients without considering the operational information at other facilities can yield worse operational outcomes for both patients and the healthcare facilities across the network , such as increased wait times for the diverted patient, and increased crowding at healthcare facilities in the network \citep{deo2011centralized,li2019review}. This has been further emphasized in the recently published literature, wherein \citep{shi2021timing} find that communication and coordination among healthcare facilities are crucial to relieve congestion across all facilities in a healthcare network, and diversion decisions must be made using information available from electronic management systems networked across all facilities under consideration \citep{adjerid2018reducing,dong2019impact,piermarini2021simulation}. In this context, our study also proposes a diversion mechanism for outpatients that implicitly assumes the presence of a centralized IT system that maintains information regarding the operational status - in terms of system state variables such as the number of patients waiting and elapsed service times in various queues within the facility - that is available to all facilities within the network.

We note here that most studies we surveyed, as is natural, considered diversion of emergency patients. Only \cite{nezamoddini2016modeling} briefly discuss the effect of diverting non-emergency patients to other healthcare facilities in a multi-hospital setting. To the best of our knowledge, we did not identify other studies that considered diversion of outpatients, and in particular, we did not identify other studies that utilized real-time LOS estimates for this purpose.  

As discussed above, our approach - in relation with the extant literature - also considers a centralized diversion mechanism involving information sharing. In this regard, our work is similar to the work by \cite{fatma2021patient} ; however, where they use real-time delay estimates to divert emergency (childbirth) patients arriving at facilities within a healthcare network seeking care, we utilize real-time delay estimates in generating LOS predictions - at the point in time when they may access the IT system associated with the healthcare network - to assign patients to a facility within the network. While this may seem like a facility assignment exercise instead of patient diversion, we consider this a diversion issue because patients are typically already assigned as belonging to the catchment area of their nearest PHC, and hence our framework determines whether they must be `diverted' to visit a facility located farther from them on the basis of real-time LOS predictions. However, where such facility assignment is not \textit{a priori} performed, our approach can be used for real-time facility assignment as well.

\section{Primary Health Centers}
\label{sec:PHC}

In India, PHCs serve as the first point of contact between the population and the healthcare provider with the objective of making healthcare services accessible in both the rural and urban areas. PHCs are established to cover a population of 30,000  in rural areas and 20,000 in hilly, tribal, and desert areas, providing a range of essential outpatient care, and limited inpatients and childbirth care \citep{Guidelines2012}.

PHCs provide six hours of outpatient care services and 24$\times$7 emergency services including inpatient and childbirth care to the patients. PHCs have one to two doctors, a non-communicable disease (NCD) nurse for conducting lifestyle diseases checks, one staff nurse per shift, a pharmacist, and a laboratory technician. Routine blood and urine investigations, and sputum tests for suspected tuberculosis cases are offered by the in-house laboratory within the PHCs. PHCs also have one labor bed for normal and assisted deliveries for childbirth patients and four to six indoor beds for admitting emergency patients to ensure easy access to public health facilities. We refer readers to \cite{shoaib2021simulation} for a detailed understanding of medical care operations at PHCs.

\begin{figure*}[htbp]
	\centering
	\includegraphics[width=15cm]{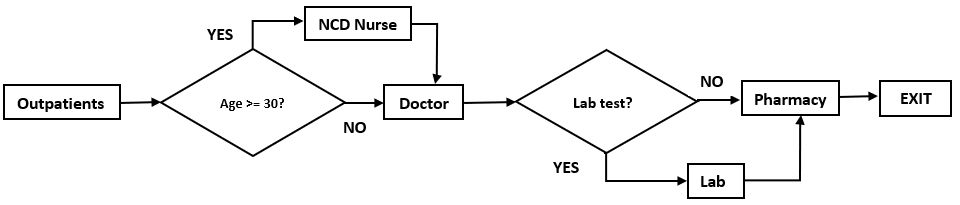}
	\caption{Outpatient flow at the PHC. Lab = laboratory.}
	\label{patflow}
\end{figure*}

We now briefly describe the patient flow within PHCs. Outpatients aged greater than 30 years first consult the NCD nurse before consulting the doctor. The NCD nurse checks blood pressure and blood glucose levels, body temperature, and also provides diet-related consultation to patients, as required. Other outpatients directly consult the doctor or wait in the outpatient queue if the doctor is busy with other patients. Once the outpatients complete their consultation with the doctor, approximately half the patients are sent to the in-house laboratory for routine investigations. All patients exit from PHC via the pharmacy regardless of whether they require medication, because the pharmacy also performs the administrative task of registering patient visits before their exit from the PHC. Outpatient flow through the PHC is depicted in Figure~\ref{patflow}.

During outpatient department (OPD) hours, inpatients and childbirth patients are attended to by the doctor first and if the doctor is unavailable, they are attended to by the staff nurse. Once their consultation with the doctor is complete, inpatients are admitted to the inpatient department (IPD) where the staff nurse monitors their condition and treats accordingly. Inpatient LOSs in the IPD range from four to six hours, because patients are typically admitted for minor conditions that require monitoring and treatment for brief periods. Childbirth patients are taken to the labour room for delivery under the supervision of the doctor and the staff nurse. After delivery, childbirth patients are shifted to the IPD ward. Note that for more complex inpatient and childbirth cases requiring more intensive and specialized care, patients are referred to higher level facilities. Outside OPD hours, patients are attended to by the staff nurse alone with doctors being summoned on very rare occasions.

As documented in \cite{shoaib2021simulation}, input parameters such as the average number of arrivals of each type of patient per day (arrival distributions were assumed to be Poisson), and the service time distribution parameters of the doctor, nurses, laboratory technicians, and the pharmacist were estimated based on data collection from in person visits to multiple PHCs. For example, the best-fit distribution for the NCD nurse service time was found to be the uniform distribution with parameters $U(2,5)$. Truncated Gaussian distributions were found to fit the service times of other resources - the doctor, laboratory technician, and pharmacist - best, with parameters $N(0.87, 0.21^2)$, $N(3.451, 0.873^2)$, and $N(2.084, 0.723^2)$ minutes, respectively. We refer readers to \cite{shoaib2021simulation} for a complete description of the parameter estimation process for other patient types served by PHCs.

The focus of this analysis is outpatient flow in the PHCs given that we consider outpatient diversion based on LOS estimates for the PHCs. We approximate the outpatient flow through the PHC operations as the flow of an entity through one more M/G/1 queueing systems in series. This approximation is possible due to the very large disparity in the arrival rates between outpatients and patients requiring admission (i.e., inpatients and childbirth patients). This is discussed in further detail in \citep{shoaib2021simulation}.

Based on the outpatient flow described in Figure~\ref{patflow} and input parameter estimates, we develop a discrete-event based simulation model of the patient care operations at two PHCs. The programming was done in Python using the \textit{Salabim} package on an Intel \textit{i}7 64-bit Microsoft Windows operating system with 16 gigabytes of memory.\textit{Salabim} is a discrete event simulation package in python developed by \cite{van2018salabim}. The package provides animation, queues, states, monitors for data collection and presentation, and simulation event tracing functionality. \textit{Salabim} has applications in transportation research, manufacturing, mining, hospital logistics, network analysis, etc. A single replication of the simulation involved a run-time horizon of 365 days with a warm-up period of 180 days. Each replication required approximately 4 minutes of computational run-time. We collect data from 40 replications and report mean and standard deviations of key performance measures at both PHCs in Table~\ref{tab1}.

\begin{table}[htbp]
	\hspace{4.5em}
	\begin{threeparttable}
		\centering
		\caption{PHC simulation outcomes}
		\begin{tabular}{|p{9.0em}|p{9.0em}|p{9.0em}|}
			\hline
              Outcomes* & PHC$_{1}$$^{+}$ & PHC$_{2}^{+}$\\ 
               \multicolumn{1}{|r|}{} & (9/1/1/1/1) & (2/1/1/1/1) \\
              \hline
                  $\rho_{doc}$ & 0.392 (0.003)& 0.935 (0.003) \\
                  \hline
                  $\rho_{NCD}$ & 0.512 (0.006) & 1.475 (0.019) \\
                  \hline
                  $\rho_{phar}$ & 0.383 (0.003) & 1.705 (0.007) \\
                	\hline
                  $\rho_{lab}$ & 0.317 (0.005) & 1.417 (0.019) \\
                \hline
                  $w_{opd}$ & 0.036 (0.001) & 0.235 (0.002) \\
                \hline
                  $w_{phar}$ & 0.243 (0.005) & 27.93 (0.618) \\
                \hline
                  $w_{lab}$ & 0.593 (0.012) & 17.47 (0.419) \\
                \hline
                 $w_{NCD}$ & 0.629 (0.023) & 30.12 (0.830) \\
                \hline
                  $LOS$ & 7.180 (0.033) & 57.31 (1.154) \\
                \hline
		\end{tabular}
            \begin{tablenotes}[para,flushleft]
        	\small
        	\item $^{+}$Outpatient interarrival time (in minutes)/number of doctors/number of NCD nurse/number of laboratory technicians/number of pharmacists.
			\item $^{*}$Resource utilizations ($\rho$) are dimensionless fractions, wait times $(w)$, and average lengths of stay $(LOS)$ are in minutes.\\
			\item $^{*}$$\rho_{doc}$: doctor's utilization, $\rho_{NCD}$: NCD nurse's utilization, $\rho_{phar}$: pharmacist's utilization, $\rho_{lab}$: laboratory technician's utilization, $w_{opd}$: OPD queue wait time, $w_{phar}$: pharmacy queue wait time, $w_{lab}$: laboratory queue wait time, $w_{NCD}$: NCD nurse queue wait time. \\
        \end{tablenotes}
          \label{tab1}%
	\end{threeparttable}
\end{table}%

From Table~\ref{tab1}, we observe that the average patient LOS at PHC$_{2}$ is significantly higher than that of PHC$_{1}$. We also note a similar trend in other performance measures such as resource utilizations and average wait time for all the staff involved in providing medical care to outpatients at both PHCs. We also see that the individual wait times at each queueing subsystem within PHC$_2$ do not appear to be prohibitively long in themselves; however, as the overall LOS in the facility is considerable given the significantly lower service times (for example, the average consultation time with the doctor is less than one minute) at these individual subsystems within the PHC. These observations motivated the development of a diversion mechanism for outpatients based on LOS predictions, which we now describe.

\section{Outpatient Diversion Mechanism}

The diversion mechanism for outpatients that we develop is depicted in Algorithm~\ref{alg:the_alg}. We describe this mechanism using the case of a specific patient  whose assigned PHC is PHC$_1$. We consider one other PHC, which we call as PHC$_2$ (located farther from the patient than PHC$_1$), as a potential destination for diversion; however, our approach can be extended to any number of facilities in a healthcare network. As discussed in Section 1, we assume that the patient can access an IT system - either a web-based or smartphone application - that communicates the LOS predictions at PHC$_1$ and PHC$_2$, and assigns the appropriate PHC to the patient based on these predictions. We assume that the patient accesses the application at time $t$, implying that the LOS prediction must be generated at time $t$; however, the LOS at each PHC must be estimated at the time the patient is likely to arrive at the facility in question. Thus, while the prediction is made at time $t$, the LOS must be estimated at time $t + \delta_1$ or $t + \delta_2$, where $\delta_1$ and $\delta_2$ are the travel times of the patient to PHC$_1$ and PHC$_2$. If we denote these LOS estimates as $LOS_{t+\delta_1}$ and $LOS_{t+\delta_2}$, then the patient is assigned the facility with the smaller of $LOS_{t+\delta_1}$ and $LOS_{t+\delta_2}$.  

\begin{algorithm}
	\caption{Outpatient diversion mechanism.}
	\label{alg:the_alg}
	\begin{algorithmic}[1]
		\State At time $t$, estimate $LOS_{t + \delta_1}$ at PHC$_1$
		\State At time $t$, estimate $LOS_{t + \delta_2}$ at PHC$_2$
		\State Visit PHC with LOS = $\min\{L_{t + \delta_1}, L_{t + \delta_2}\}$
	\end{algorithmic}
\end{algorithm}

We note here that for the purposes of this study, we assume that the patient is required to visit the assigned PHC, and incurs some penalty if they visit any other facility in the network; however, this mechanism, if adopted, is realistically more likely to provide a choice to the patient. In such situations the probability of compliance with the diversion or facility assignment suggestion must be considered in the analysis. We do not explore this in our study, and leave it as an immediate avenue of future research.

Another consideration here is that the diversion decision can be based on whether the sum of travel time to the PHC under consideration plus predicted LOS at the PHC is less than the corresponding quantity at the other PHCs in the network. We do not consider this case in our current version of the LOS-based diversion mechanism; this is because we wish to introduce this real-time LOS prediction based diversion mechanism in its simplest form (i.e., based only on predicted LOS at each facility). However, the diversion mechanism in Algorithm~\ref{alg:the_alg} can easily be modified to take such additional considerations into account, and we intend to explore this specific consideration in future work. Note also that the estimates of travel times used in the above algorithm can be considered to be average values; however, the actual values used in the diversion algorithm can be real-time estimates obtained from applications such as Google Maps depending upon the application process interface links that can be integrated into the IT system used for deployment of the diversion mechanism.

It is evident that the key steps within the diversion mechanism in Algorithm~\ref{alg:the_alg} are the estimation of the LOSs $L_{t + \delta_1}$ and $L_{t + \delta_2}$. We now describe how these LOS predictions are generated.

\subsection{Real-time Length of Stay Prediction}
\label{lospred}

As discussed in Section~\ref{sec:PHC}, we approximate the outpatient care process at the PHC as receiving care from one or more M/G/1 systems in series. The four possible M/G/1 system pathways for an outpatient in the PHC have also been depicted in Figure~\ref{patflow}. We estimate the LOS in the PHC as the sum of the LOSs in each of the M/G/1 subsystems where the patient is likely to receive service. 

\begin{equation}
	\label{eq:lostot}
	LOS_{phc} = LOS_{doc} + LOS_{NCD} + LOS_{lab} + LOS_{pharmacy}
\end{equation}

The terms in equation~\ref{eq:lostot} are self-explanatory. We note here that for $LOS_{phc}$ to be as accurate as possible, the LOSs associated with the laboratory and NCD nurse queueing systems must be weighted with the probabilities that the patient avails of service from these subsystems. However, we do not do this because of precedence in the literature regarding providing quantiles higher than the 50\textsuperscript{th} quantile of the distribution of the predicted delay - or in this case, the predicted LOS - to increase customer satisfaction in queues \citep{whitt1999predicting,ibrahim2018sharing}. Note that this implies that we overestimate the LOS; however, given that each of the terms in equation~\ref{eq:lostot} scales in the same manner with patient load, and does so to the same extent for all facilities considered, the diversion decision itself is not affected by this overestimation. Each LOS term in the right-hand side of equation~\ref{eq:lostot} is estimated as the sum of the prediction of the delay experienced by the patient at the M/G/1 system under question and the average service time at the system. This is given in equation~\ref{eq:lossub} below.

\begin{equation}
	\label{eq:lossub}
	LOS_{s} = d_{s} + E[s], \forall~s \in S 
\end{equation}

In equation~\ref{eq:lossub}, $s$ is a subscript indicating an M/G/1 queueing subsystem in the PHC, and $s \in S = \{doc, NCD, lab, pharmacy\}$. $d_{s}$ represents the estimated delay for the patient at a given M/G/1 subsystem, and $E[s]$ represents the expected value of the service time. The estimated delay $d_s$ is in turn estimated as a function of the number of patients ahead of the patient under consideration in the queue, and the expected remaining service time for the patient currently receiving service at the queueing system. This is expressed below. 

\begin{equation}
	\label{eq:wtpred}
	d_{s} = L_{q(s)} E[s] + r_e,~ \forall~s \in S
\end{equation}

In equation~\ref{eq:wtpred}, $L_{q(s)}$ represents the number of patients ahead of the patient under consideration in the queue at subsystem $s$, and $r_e$ represents the prediction of the remaining service time given that the elapsed service time for the patient receiving service is $x$. We now discuss how $r_e$ is generated. For M/G/1 systems, the estimation of $r_e$ is traditionally done as in equation~\ref{eq:wgen}. In equation~\ref{eq:wgen}, $T$ is the random variable representing the service time ($t$ is thus its realization), $G$ is the $cdf$ of the service time.

\begin{equation}
	\label{eq:wgen}
	\begin{aligned}
		&P(T \leq t | x) = \frac{P(x \leq T \leq t + x)}{P(T \geq x)}\\
		& \implies G(t|x) =  \frac{G(t + x) - G(x)}{1-G(x)}
	\end{aligned}
\end{equation}

$r_e$ is then estimated as the expected value of the remaining service time by calculating the $pdf$ $g(t|X)$ from equation~\ref{eq:wgen}. However, performing this calculation can prove tedious depending upon the nature of the distribution $G(t|x)$; this is discussed in detail in \cite{fatma21}. For example, computing $r_e$ via equation~\ref{eq:wgen} for the triangular distribution requires working with a piecewise continuous $cdf$, and for the Gaussian distribution requires numerical computation of the integrals involved in computation of $r_e$ (which involve the Gaussian error function). Therefore, we propose an alternate delay predictor for such systems, which can potentially also be extended to multi-class multi-server queueing systems, that is considerably easier to compute and implement. This delay predictor is given below. 

\begin{equation}
	\label{eqwdel}
	r_e =
	\begin{cases}
		G^{-1}{(0.5)}-x ,        & \text{$ 0\le x < G^{-1}{(0.5)}$} \\
		G^{-1}{(0.75)}-x,       & \text{$ G^{-1}{(0.5)}\le x< G^{-1}{(0.75)}$} \\
		\frac{(b-x)}{2},          & \text{$G^{-1}{(0.75)}\le x\le b$} \\
		
	\end{cases}
\end{equation}

Here, $r_e$ is estimated depending upon the range in which the elapsed service time $x$ of the patient currently in service lies with respect to the service time distribution. $G^{-1}{(0.5)}$, and $G^{-1}{(0.75)}$ are 50$^{th}$ and 75$^{th}$ quantile of the service time distribution and $b$ is the extreme quantile (e.g., equal to $G^{-1}(0.99)$) or the upper limit of the service time  $cdf$ (e.g., where the $cdf$ has bounded support). 

$r_e$ as estimated from equation~\ref{eqwdel} can then be used in conjunction with equations~\ref{eq:wtpred} and~\ref{eq:lossub} to estimate the LOS at the subsystem under consideration. This computation would suffice if the LOS prediction is required to be made at time $t$; however, we require LOS predictions at time $t + \delta$, where $\delta$ is the time required for the patient to arrive at the PHC. LOS predictions at time $t+\delta$ are generated as follows.

In order to calculate the LOS prediction at time $t + \delta$, we must first estimate the effective average queue length at $t + \delta$, $L_{q(t + \delta)}$, as a function of the queue length at time $t$, given by $L_{q(t)}$. This can be accomplished using the average post-diversion patient arrival and service rates, $\lambda_{e}$ and $E[s]$, respectively. This is given below. 

\begin{equation}
	\label{eq:lqdelta}
	L_{q(t+\delta)} = L_q + (\lambda_{e} - 1) - \left \lfloor{\max\left\{ \frac{\delta - r_{e(t)}}{E[s]}, 0 \right\}}\right \rfloor
\end{equation}

In equation~\ref{eq:lqdelta}, we substract 1 from $\lambda_{e}$ to account for the patient under consideration for diversion, and the third term on the right hand side represents the average number of patients that can be served in $\delta$ time units after taking into account the remaining service time $r_{e(t)}$ of the patient in service at time $t$. The average elapsed service time at time $t + \delta$, denoted by $x_{t + \delta}$, is thus estimated as the remainder of $\frac{\delta - r_{e(t)}}{E[s]}$, and the average remaining service time at $t + \delta$, denoted by $w_{e(t + \delta)}$, is estimated by inputting $x_{t + \delta}$ into equation~\ref{eqwdel}. The expected delay at time $t + \delta$ for a patient arriving to subsystem $s$, $d_{s(t + \delta)}$, can thus be estimated by inputting $L_{q,s(t+\delta)}$ and $r_{e,s(t+\delta)}$ into equation~\ref{eq:wtpred}, and the LOS at time $t+\delta$ can in turn be generated using the value of $d_{s(t + \delta)}$ in equation~\ref{eq:lossub}. This is summarized in equation~\ref{eq:finldelta} below.

\begin{equation}
	\label{eq:finldelta}
	\begin{aligned}
	&LOS_{s(t + \delta)}  = d_{s(t+\delta)} + E[s]\\
	&\text{where } d_{s(t+\delta)} = L_{q,s(t+\delta)} E[s] + r_{e,s(t+\delta)} 
	\end{aligned}
\end{equation}

The LOS at subsystem $s$ estimated from equation~\ref{eq:finldelta} can be used to calculate the LOS at each of the subsystems in $S$, and then summed per equation~\ref{eq:lostot} to yield the total LOS at the PHC. Note that the LOS prediction at each subsystem must be calculated at the point at which the predicted LOS at the previous subsystem ends. This process of successively calculating the LOSs is summarized in Algorithm~\ref{alg:loscalc}

\begin{algorithm}
	\caption{Calculation of the total LOS at a PHC.}
	\label{alg:loscalc}
	\begin{algorithmic}[1]
		\State Initialize with the ordered tuple $S_{ord} = (NCD, doc, lab, pharmacy)$, $\delta$ = travel time to the PHC, current time $t$, $LOS_{tot} = 0$. 
		\For{$s \in S_{ord}$}
		\State Estimate LOS at $s$ at time $t + \delta$ as $LOS_{s(t + \delta)}$.
		\State  $LOS_{tot} = LOS_{tot} + LOS_{s(t + \delta)}$. 
		\State $\delta = \delta + LOS_{s(t + \delta)}$
		\EndFor
	\end{algorithmic}
\end{algorithm}

We must mention here that in the implementation of the above algorithm, the arrival rate at each subsystem is assumed to be the same - that is, $\lambda_e$. This is possible only because we assume that patients visit all subsystems in series \citep{hopp2011factory}, and we consider relaxing this assumption as an immediate avenue of future work. We describe the outcomes of implementing the diversion mechanism in the following section. 

We also note here that the deployment and generation of these LOS predictions is unlikely to incur significant computational expense, regardless of the size of the facility network. This is because the generation of the LOS predictions at a facility involves just a single function evaluation, and this will remain the same regardless of the type of LOS predictor involved - for example, a queueing theory based predictor or a statistical learning based predictor. However, the specific LOS predictor to be employed must be selected taking into account the number of system state variables that are required for LOS prediction. For example, if a particular LOS predictor requires a large number of system state variables to be recorded and updated at high frequency, then the complexity of the IT system required for this purpose will also increase. On the other hand, if another LOS predictor is available that requires significantly fewer variables to be recorded at a small loss of predictor accuracy, then it might be prudent to choose this latter LOS predictor from the standpoint of ease of deployment.

Now, based on the real-time LOS estimates generated using Algorithm~\ref{alg:loscalc}, we can implement the diversion mechanism in Algorithm~\ref{alg:the_alg}.

\section{Simulation Experiments}
\label{divres}
In this section, we present results from implementing the diversion mechanism described in Algorithm~\ref{alg:the_alg}. We compare the operational outcomes at both PHCs for three cases: (a) without diversion, (b) diversion implemented using actual real-time LOS estimates (which are available to the simulation modeler), and (c) the real-time LOS predictions generated using the methodology described in Section~\ref{lospred}. Before we present the operational outcomes, we discuss the accuracy of the real-time LOS predictor. 

We characterize the accuracy of the LOS predictor using the mean absolute percentage error (MAPE) score, which is given by the formula $\text{MAPE} = \frac{1}{N}\sum\limits_{i=1}^{N}\left|\frac{LOS_{a,i}-LOS_{p,i}}{LOS_{a,i}}\right|$. Here {LOS}$_{a,i}$ is the actual LOS estimate of the i$^{th}$ patient and {LOS}$_{p,i}$ is the predicted LOS estimate of the i$^{th}$ patient and $N$ is the number of patients in the sample. When LOS predictions are generated for the two PHCs that we consider (PHC$_{1}$ and PHC$_{2}$) in the first set of computational experiments that we describe in the following paragraphs, we find that the MAPE scores are 13.00\% and 18.14\% at PHC$_{1}$ and PHC$_{2}$, respectively. While this appears to indicate reasonable performance of the LOS predictor, several avenues of improvement of the predictor remain, which we discuss in the following section.

In Table~\ref{tab2}, we report results from the implementation of the diversion mechanism for PHC$_{1}$ and PHC$_{2}$ with operational parameters corresponding to those described in Table~\ref{tab1}. Given that diversion is likely to make the utilization of resources more equitable across a network of facilities, and due to space limitations, we report the mean and standard deviation of the percentage difference between operational outcomes such as resource utilizations, wait time, and average LOSs across both healthcare facilities in Table~\ref{tab2}. We also report the percentage of patients who are diverted - that is, those who visit a PHC than their assigned PHC. We report these outcomes for the three cases discussed in the above paragraphs.

\begin{table}[htbp]
	\hspace{2.5em}
	\begin{threeparttable}
		\centering
		\caption{Outpatient diversion: operational outcomes across both PHCs.}
		\begin{tabular}{|p{5.25em}|p{6.0em}|p{7.5em}|p{9.0em}|}
			\hline
			Outcomes & No diversion & With actual LOS  & With predicted LOS \\
			\hline
			$\Delta_{\rho_{doc}}$ & 58.09 (0.36) & 5.91 (0.03) & 15.60 (0.01) \\
			\hline
			$\Delta_{\rho_{ncd}}$ & 65.26 (0.58) & 7.82 (0.02) & 13.50 (0.02) \\
			\hline
			$\Delta_{\rho_{phar}}$ & 77.52 (0.18) & 16.33 (0.02) & 22.20 (0.04) \\
			\hline
			$\Delta_{\rho_{lab}}$ & 77.59 (0.45) & 10.30 (0.03) & 18.02 (0.06) \\
			\hline
			$\Delta_{w_{opd}}$& 84.53 (0.38) & 4.07 (0.02) & 46.77 (0.03) \\
			\hline
			$\Delta_{w_{phar}}$& 99.12 (0.02) & 51.55 (0.08) & 74.28 (0.03) \\
			\hline
			$\Delta_{w_{lab}}$ & 96.60 (0.09) & 37.95 (0.05) & 79.03 (0.04) \\
			\hline
			$\Delta_{w_{ncd}}$ & 97.91 (0.10) & 14.12 (0.10) & 86.16 (0.02) \\
			\hline
			$\Delta_{LOS}$ & 87.46 (0.26) & 10.21 (0.04) & 63.41 (0.08) \\
			\hline
			$\beta^{*}$ & 0.00 (0.00) & 30.95 (2.98) & 45.25 (1.48) \\
			\hline
		\end{tabular}
		\begin{tablenotes}[para,flushleft]
			\tiny
			\item $\Delta_{\rho_{doc}}$ = difference in doctor's utilization, $\Delta_{\rho_{NCD}}$= difference in NCD nurse utilization, $\Delta_{\rho_{phar}}$= difference in pharmacist's utilization, $\Delta_{\rho_{lab}}$= difference in laboratory technician's utilization, $\Delta_{w_{opd}}$ = difference in OPD queue wait time, $\Delta_{w_{phar}}$ = difference in pharmacy queue wait time, $\Delta_{w_{lab}}$ = difference in laboratory queue wait time, $\Delta_{w_{NCD}}$ = difference in NCD nurse queue wait time, $\beta^{*}$ = proportion (\%) of outpatients who are diverted.\\
		\end{tablenotes}
		\label{tab2}%
	\end{threeparttable}
\end{table}%

It is evident from Table~\ref{tab2} that implementing diversion leads to significantly more equitable operational outcomes across both PHCs. Without diversion, as seen from Table~\ref{tab1}, the LOS at PHC$_2$ was 57.31 minutes in comparison to an LOS of 7.18 minutes at PHC$_1$. When diversion is implemented using the actual real-time LOS estimates, we see that the LOS at PHC$_2$ reduces to 15.33 minutes whereas the LOS at PHC$_1$ increases to 13.75 minutes. With diversion implemented using the real-time LOS predictor, we see LOSs of 25.92 minutes at PHC$_1$ and 9.50 minutes at PHC$_2$. Thus it is clear that diversion, even using the real-time LOS predictor, yields a reduction in LOS at the more heavily utilized system.

From Table~\ref{tab2}, we see similar trends in the other outcomes as well when diversion is implemented using the real-time LOS predictor. Overall, we see a significant decrease in the congestion in PHC$_2$. A key observation is that the extent to which operational outcomes become equitable across both PHCs depend on the accuracy of the delay predictor employed - that is, the relative difference in a resource's utilization or the wait time at a subsystem increases with a decrease in the accuracy of the delay predictor used. Another important observation is that the proportion of patients diverted also increase with decrease in delay predictor accuracy. However, this is likely due to the significant difference in the outpatient loads between the two PHCs prior to implementing diversion - that is, an average interarrival time of 9 minutes at PHC$_1$ compared to an average interarrival time of 2 minutes at PHC$_2$.

In order to check how the operational outcomes change with diversion when the difference in outpatient loads across both PHCs is not as high as in the previous case, we conducted a sensitivity analysis by changing the average interarrival times of outpatients to 2 minutes at PHC$_{1}$ and 4 minutes at PHC$_{2}$. We present the results from this sensitivity analysis results in Table~\ref{tab3} and observe similar trends in the operational outcomes across both PHCs as observed in Table~\ref{tab2}. We note that the MAPE scores of the real-time LOS predictor that we propose at PHC$_{1}$ and PHC$_{2}$ were found to be 13.82\% and 18.33\% respectively, respectively.

\begin{table}[htbp]
	\hspace{2.5em}
	\begin{threeparttable}
		\centering
		\caption{Sensitivity analysis: operational outcomes across both PHCs.}
		\begin{tabular}{|p{5.25em}|p{6.0em}|p{7.5em}|p{9.0em}|}
			\hline
			Outcomes & No diversion & With actual LOS & With predicted LOS \\
			\hline
			$\Delta_{\rho_{doc}}$ & 37.28 (0.41) & 18.82 (0.05) & 24.91 (0.01) \\
			\hline
			$\Delta_{\rho_{ncd}}$ & 41.92 (0.85) & 19.84 (0.04) & 26.68 (0.02) \\
			\hline
			$\Delta_{\rho_{phar}}$ & 49.92 (0.85) & 22.96 (0.04) & 34.08 (0.01) \\
			\hline
			$\Delta_{\rho_{lab}}$ & 50.02 (1.07) & 24.28 (0.04) & 33.29 (0.02) \\
			\hline
			$\Delta_{w_{opd}}$& 60.01 (0.84) & 11.52 (0.05) & 17.13 (0.07) \\
			\hline
			$\Delta_{w_{phar}}$& 96.19 (0.08) & 46.19 (0.07) & 64.84 (0.03) \\
			\hline
			$\Delta_{w_{lab}}$ & 87.59 (0.52) & 44.81 (0.09) & 71.24 (0.03) \\
			\hline
			$\Delta_{w_{ncd}}$ & 92.52 (0.23) & 29.18 (0.07) & 71.24 (0.03) \\
			\hline
			$\Delta_{LOS}$ & 83.15 (0.29) & 31.48 (0.07) & 78.96 (0.01) \\
			\hline
			$\beta^{*}$ & 0.00 (0.00) & 45.37 (0.78) & 36.69 (0.10) \\
			\hline
		\end{tabular}
		\label{tab3}%
		\begin{tablenotes}[para,flushleft]
			\tiny
			\item $\Delta_{\rho_{doc}}$ = difference in doctor's utilization, $\Delta_{\rho_{NCD}}$= difference in NCD nurse utilization, $\Delta_{\rho_{phar}}$= difference in pharmacist's utilization, $\Delta_{\rho_{lab}}$= difference in laboratory technician's utilization, $\Delta_{w_{opd}}$ = difference in OPD queue wait time, $\Delta_{w_{phar}}$ = difference in pharmacy queue wait time, $\Delta_{w_{lab}}$ = difference in laboratory queue wait time, $\Delta_{w_{NCD}}$ = difference in NCD nurse queue wait time, $\beta^{*}$ = proportion (\%) of outpatients who are diverted.\\
		\end{tablenotes}
	\end{threeparttable}
\end{table}

Once again, as expected, we notice that the extent to which operational outcomes become equitable across both PHCs decreases with a decrease in the accuracy of the LOS predictor employed. However, when compared to the corresponding outcome in Table~\ref{tab2}, we notice the proportion of patients diverted actually decreases with a decrease in the accuracy of the LOS predictor. That is, in this case, the proportion diverted decreases with the less accurate LOS predictor. Thus, our results indicate that the proportion of patients diverted across a healthcare facility network, when diverted on the basis of real-time LOS predictions, depend on both the accuracy of LOS predictor employed as well as the differences in patient loads across the facilities within the network.

Overall, it is evident that outpatient diversion using real-time LOS predictions appears to improve operational performance of a healthcare facility network from both the provider as well as patient standpoints. 

We now conclude our work with a discussion of the contributions of our work and avenues for further research.

\section{\uppercase{Conclusion}}
\label{sec:conclusion}

In this work, we present preliminary results from the simulated implementation of a framework for outpatient diversion using real-time LOS predictions. We emphasize here that the real-time LOS predictions and the diversion decision are made at the point of origin of the patient, as opposed to when the patient arrives at a healthcare facility. We empirically show via discrete-event simulation that patient diversion results in more equitable utilization of resources across the healthcare facilities that we consider when compared to the case without diversion. The diversion framework that we propose can also be repurposed for real-time facility assignment in health systems where patients are not \textit{a priori} assigned their first point of contact for medical care.

To the best of our knowledge, our study is the first of its kind that considers real-time LOS predictions in initiating diversion across healthcare facilities. Previous studies initiated diversion using factors such as average wait time, number of beds occupied, number of patients in waiting queue. Further, we demonstrate how LOS can become a criterion for diversion especially when diversion of non-emergency patients, and in particular outpatients, is considered. Once again, we did not come across a previously published study that considered outpatient diversion. Finally, to our knowledge, our study is also the first to develop a methodology for the real-time estimation of LOSs in the near future. This in turn utilizes a novel approximation of the real-time delay predictor for M/G/1 queueing systems. In comparison to most studies in the literature that utilize LOS predictors based on statistical learning methods, our analytical approach grounded in queueing theory may be easier to program and deploy in practice. Further, our approach also opens up avenues for leveraging the substantial body of work regarding real-time delay prediction for real-time LOS-based diversion for more complex queueing systems than that encountered in this study.

A key avenue of future research involves improving the accuracy of the real-time LOS predictor that we develop in this study. While the accuracy of our real-time LOS predictor appears reasonable, with MAPEs not exceeding 20\%, relaxing our assumption that all patients avail service from all subsystems within the PHC is likely to improve the accuracy of the predictor. A related direction for future work involves benchmarking our work against statistical machine learning based predictors of LOS, and assessing how the operational outcomes compare with respect to the outcomes associated with our queuing theory based predictor. Similarly, our LOS based diversion mechanism can also be compared to other diversion mechanisms, such as wait-time based mechanisms.

A key question that arises here is how the implementation of the diversion mechanism in Algorithm~\ref{alg:the_alg} would scale in terms of computational and operational overheads when the size of the network increases. One of the questions that must be answered prior to deploying this approach in a larger network would involve determining, for each facility in the network, the set of facilities at which real-time LOS predictions must be generated in order to determine the diversion decision. For example, in a network with 1,000 facilities, in order to determine the diversion decision for a patient, real-time LOS predictions need not be generated at each of the 1,000 PHCs (as many would be located at a significant distance from the patient's assigned PHC). Thus determining the optimal set of facilities that need to be considered as diversion destinations for a given facility in a large network becomes an important avenue of future research when considering the scaling of this approach to larger facility networks. As part of this, an immediate avenue of research we are pursuing includes expanding this diversion framework to the entire network of 9 PHCs in the Indian district that we consider, and we plan to answer the above question for this network.

Finally, another avenue of future research is extending our work to more complex queueing systems, including multi-class multi-server systems, which may be applicable in the case of diversion of inpatients or emergency patients.  These more complex queueing systems can include queueing systems with non-stationary arrival and service processes. Real-time delay predictors have been developed for certain queueing systems with non-stationary arrival, service or abandonment processes \cite{ibrahim2018sharing}; and LOSs for such systems may correspondingly be calculated in a manner similar to that which we have described in this paper. However, in case the non-stationarity is not complex - for example, increased arrival rates may be observed during the weekends, and arrival rates remain the same during all weekdays - then the methodology described for LOS prediction in this paper can directly be applied by just changing the arrival rates in the LOS predictor expressions depending upon the day of the week.

\bibliographystyle{apalike}
\bibliography{interactapasample}

\begin{thebibliography}{}

\bibitem[Adjerid et~al., 2018]{adjerid2018reducing}
Adjerid, I., Adler-Milstein, J., and Angst, C. (2018).
\newblock Reducing medicare spending through electronic health information
  exchange: the role of incentives and exchange maturity.
\newblock {\em Information Systems Research}, 29(2):341--361.

\bibitem[Aghajani and Kargari, 2016]{aghajani2016determining}
Aghajani, S. and Kargari, M. (2016).
\newblock Determining factors influencing length of stay and predicting length
  of stay using data mining in the general surgery department.
\newblock {\em Hospital Practices and Research}, 1(2):53--58.

\bibitem[Alsinglawi et~al., 2020]{alsinglawi2020predicting}
Alsinglawi, B., Alnajjar, F., Mubin, O., Novoa, M., Alorjani, M., Karajeh, O.,
  and Darwish, O. (2020).
\newblock Predicting length of stay for cardiovascular hospitalizations in the
  intensive care unit: Machine learning approach.
\newblock In {\em 2020 42nd Annual International Conference of the IEEE
  Engineering in Medicine \& Biology Society (EMBC)}, pages 5442--5445. IEEE.

\bibitem[Arefian et~al., 2019]{arefian2019estimating}
Arefian, H., Hagel, S., Fischer, D., Scherag, A., Brunkhorst, F.~M., Maschmann,
  J., and Hartmann, M. (2019).
\newblock Estimating extra length of stay due to healthcare-associated
  infections before and after implementation of a hospital-wide infection
  control program.
\newblock {\em PloS one}, 14(5):e0217159.

\bibitem[Ayyoubzadeh et~al., 2020]{Ayy20}
Ayyoubzadeh, S.~M., Ghazisaeedi, M., Kalhori, S. R.~N., Hassaniazad, M.,
  Baniasadi, T., Maghooli, K., and Kahnouji, K. (2020).
\newblock A study of factors related to patients’ length of stay using data
  mining techniques in a general hospital in southern iran.
\newblock {\em Health information science and systems}, 8(1):1--11.

\bibitem[Baek et~al., 2018]{baek2018analysis}
Baek, H., Cho, M., Kim, S., Hwang, H., Song, M., and Yoo, S. (2018).
\newblock Analysis of length of hospital stay using electronic health records:
  A statistical and data mining approach.
\newblock {\em PloS one}, 13(4):e0195901.

\bibitem[Barnes et~al., 2016]{barnes2016real}
Barnes, S., Hamrock, E., Toerper, M., Siddiqui, S., and Levin, S. (2016).
\newblock Real-time prediction of inpatient length of stay for discharge
  prioritization.
\newblock {\em Journal of the American Medical Informatics Association},
  23(e1):e2--e10.

\bibitem[Blanchard et~al., 2021]{blanchard2021vision}
Blanchard, J., Washington, R., Becker, M., Vasanthakumar, N., Gopal, K.~M., and
  Sarwal, R. (2021).
\newblock Vision 2035 public health surveillance in india.

\bibitem[Deo and Gurvich, 2011]{deo2011centralized}
Deo, S. and Gurvich, I. (2011).
\newblock Centralized vs. decentralized ambulance diversion: A network
  perspective.
\newblock {\em Management Science}, 57(7):1300--1319.

\bibitem[Dong et~al., 2019]{dong2019impact}
Dong, J., Yom-Tov, E., and Yom-Tov, G.~B. (2019).
\newblock The impact of delay announcements on hospital network coordination
  and waiting times.
\newblock {\em Management Science}, 65(5):1969--1994.

\bibitem[El-Bouri et~al., 2021]{el2021machine}
El-Bouri, R., Taylor, T., Youssef, A., Zhu, T., and Clifton, D.~A. (2021).
\newblock Machine learning in patient flow: a review.
\newblock {\em Progress in Biomedical Engineering}.

\bibitem[Fatma et~al., 2020]{fatma2020primary}
Fatma, N., Mohd, S., Ramamohan, V., and Mustafee, N. (2020).
\newblock Primary healthcare delivery network simulation using stochastic
  metamodels.
\newblock In {\em 2020 Winter Simulation Conference (WSC)}, pages 818--829.
  IEEE.

\bibitem[Fatma and Ramamohan, 2021a]{fatma2021patient}
Fatma, N. and Ramamohan, V. (2021a).
\newblock Patient diversion across primary health centers using real time delay
  predictors.
\newblock In {\em 2021 Institute of Industrial and Systems Engineers (IISE)
  Annual Conference \& Expo}.

\bibitem[Fatma and Ramamohan, 2021b]{fatma21}
Fatma, N. and Ramamohan, V. (2021b).
\newblock Patient diversion using real-time delay prediction across healthcare
  facility networks.
\newblock \url{http://web.iitd.ac.in/~mez188287/NF.pdf}, accessed
  29\textsuperscript{th} November 2021.

\bibitem[Guti{\'e}rrez et~al., 2021]{gutierrez2021predicting}
Guti{\'e}rrez, J. M.~P., Sicilia, M.-{\'A}., Sanchez-Alonso, S., and
  Garc{\'\i}a-Barriocanal, E. (2021).
\newblock Predicting length of stay across hospital departments.
\newblock {\em IEEE Access}, 9:44671--44680.

\bibitem[Hijry and Olawoyin, 2020]{hijry2020application}
Hijry, H. and Olawoyin, R. (2020).
\newblock Application of machine learning algorithms for patient length of stay
  prediction in emergency department during hajj.
\newblock In {\em 2020 IEEE International Conference on Prognostics and Health
  Management (ICPHM)}, pages 1--8. IEEE.

\bibitem[Hopp and Spearman, 2011]{hopp2011factory}
Hopp, W.~J. and Spearman, M.~L. (2011).
\newblock {\em Factory physics}.
\newblock Waveland Press.

\bibitem[Ibrahim, 2018]{ibrahim2018sharing}
Ibrahim, R. (2018).
\newblock Sharing delay information in service systems: a literature survey.
\newblock {\em Queueing Systems}, 89(1):49--79.

\bibitem[IPHS-Guidelines, 2012]{Guidelines2012}
IPHS-Guidelines (2012).
\newblock {\em Guidelines for Primary Health Centres}.
\newblock Directorate General of Health Services, New Delhi, India.

\bibitem[Li et~al., 2019]{li2019review}
Li, M., Vanberkel, P., and Carter, A.~J. (2019).
\newblock A review on ambulance offload delay literature.
\newblock {\em Health care management science}, 22(4):658--675.

\bibitem[Liu et~al., 2018]{liu2018patients}
Liu, Y., Zhong, L., Yuan, S., and van~de Klundert, J. (2018).
\newblock Why patients prefer high-level healthcare facilities: a qualitative
  study using focus groups in rural and urban china.
\newblock {\em BMJ global health}, 3(5):e000854.

\bibitem[Marquinez et~al., 2021]{marquinez2021identifying}
Marquinez, J.~T., Saur{\'e}, A., Cataldo, A., and Ferrer, J.-C. (2021).
\newblock Identifying proactive icu patient admission, transfer and diversion
  policies in a public-private hospital network.
\newblock {\em European Journal of Operational Research}.

\bibitem[MoHFW, 2020]{phc2020}
MoHFW, I. (2020).
\newblock {\em Primary Health Centers (PHCs)}.
\newblock Ministry of Health and Family Welfare, India.
\newblock accessed 6\textsuperscript{th} October
  2021.\url{https://pib.gov.in/PressReleasePage.aspx?PRID=1656190}.

\bibitem[Morley et~al., 2018]{morley2018emergency}
Morley, C., Unwin, M., Peterson, G.~M., Stankovich, J., and Kinsman, L. (2018).
\newblock Emergency department crowding: a systematic review of causes,
  consequences and solutions.
\newblock {\em PloS one}, 13(8):e0203316.

\bibitem[Mustafee and Powell, 2020]{mustafee2020providing}
Mustafee, N. and Powell, J. (2020).
\newblock Providing real-time information for urgent care.
\newblock {\em Impact}, 2021(1):25--29.

\bibitem[Nezamoddini and Khasawneh, 2016]{nezamoddini2016modeling}
Nezamoddini, N. and Khasawneh, M.~T. (2016).
\newblock Modeling and optimization of resources in multi-emergency department
  settings with patient transfer.
\newblock {\em Operations Research for Health Care}, 10:23--34.

\bibitem[Piermarini and Roma, 2021]{piermarini2021simulation}
Piermarini, C. and Roma, M. (2021).
\newblock A simulation-based optimization approach for analyzing the ambulance
  diversion phenomenon in an emergency department network.
\newblock {\em arXiv preprint arXiv:2108.04162}.

\bibitem[Ramirez-Nafarrate et~al., 2011]{ramirez2011design}
Ramirez-Nafarrate, A., Fowler, J.~W., and Wu, T. (2011).
\newblock Design of centralized ambulance diversion policies using
  simulation-optimization.
\newblock In {\em Proceedings of the 2011 Winter Simulation Conference (WSC)},
  pages 1251--1262. IEEE.

\bibitem[Rao and Sheffel, 2018]{rao2018quality}
Rao, K.~D. and Sheffel, A. (2018).
\newblock Quality of clinical care and bypassing of primary health centers in
  india.
\newblock {\em Social science \& medicine}, 207:80--88.

\bibitem[Shaaban et~al., 2021]{shaaban2021statistical}
Shaaban, A.~N., Peleteiro, B., and Martins, M. R.~O. (2021).
\newblock Statistical models for analyzing count data: predictors of length of
  stay among hiv patients in portugal using a multilevel model.
\newblock {\em BMC Health Services Research}, 21(1):1--17.

\bibitem[Sharma et~al., 2021]{sharma2021overcrowding}
Sharma, R., Prakash, A., Chauhan, R., and Dhibar, D.~P. (2021).
\newblock Overcrowding an encumbrance for an emergency health-care system: A
  perspective of health-care providers from tertiary care center in northern
  india.
\newblock {\em Journal of Education and Health Promotion}, 10.

\bibitem[Shi et~al., 2021]{shi2021timing}
Shi, P., Helm, J.~E., Deglise-Hawkinson, J., and Pan, J. (2021).
\newblock Timing it right: Balancing inpatient congestion vs. readmission risk
  at discharge.
\newblock {\em Operations Research}.

\bibitem[Shoaib and Ramamohan, 2021]{shoaib2021simulation}
Shoaib, M. and Ramamohan, V. (2021).
\newblock Simulation modelling and analysis of primary health centre
  operations.
\newblock {\em arXiv preprint arXiv:2104.12492}.

\bibitem[Song et~al., 2015]{song2015diseconomies}
Song, H., Tucker, A.~L., and Murrell, K.~L. (2015).
\newblock The diseconomies of queue pooling: An empirical investigation of
  emergency department length of stay.
\newblock {\em Management Science}, 61(12):3032--3053.

\bibitem[Turgeman et~al., 2017]{turgeman2017insights}
Turgeman, L., May, J.~H., and Sciulli, R. (2017).
\newblock Insights from a machine learning model for predicting the hospital
  length of stay (los) at the time of admission.
\newblock {\em Expert Systems with Applications}, 78:376--385.

\bibitem[van~der Ham, 2018]{van2018salabim}
van~der Ham, R. (2018).
\newblock salabim: discrete event simulation and animation in python.
\newblock {\em Journal of Open Source Software}, 3(27):767.

\bibitem[Verburg et~al., 2014]{verburg2014comparison}
Verburg, I.~W., de~Keizer, N.~F., de~Jonge, E., and Peek, N. (2014).
\newblock Comparison of regression methods for modeling intensive care length
  of stay.
\newblock {\em PloS one}, 9(10):e109684.

\bibitem[Wachtel and Elalouf, 2020]{wachtel2020addressing}
Wachtel, G. and Elalouf, A. (2020).
\newblock Addressing overcrowding in an emergency department: an approach for
  identifying and treating influential factors and a real-life application.
\newblock {\em Israel Journal of Health Policy Research}, 9(1):1--12.

\bibitem[Whiteside et~al., 2020]{whiteside2020redesigning}
Whiteside, T., Kane, E., Aljohani, B., Alsamman, M., and Pourmand, A. (2020).
\newblock Redesigning emergency department operations amidst a viral pandemic.
\newblock {\em The American journal of emergency medicine}, 38(7):1448--1453.

\bibitem[Whitt, 1999]{whitt1999predicting}
Whitt, W. (1999).
\newblock Predicting queueing delays.
\newblock {\em Management Science}, 45(6):870--888.

\bibitem[Zhou et~al., 2019]{zhou2019estimating}
Zhou, Q., Fan, L., Lai, X., Tan, L., and Zhang, X. (2019).
\newblock Estimating extra length of stay and risk factors of mortality
  attributable to healthcare-associated infection at a chinese university
  hospital: a multi-state model.
\newblock {\em BMC infectious diseases}, 19(1):1--7.

\end{thebibliography}

\end{document}